\documentclass[12pt]{article}
\usepackage[]{natbib}
\usepackage{hyperref}
\usepackage{graphicx} 
\usepackage{url}
\usepackage[margin=1in]{geometry}

\title{Solar System Surface Technosignatures: Searching for Signs of Artificial Design on Extraterrestrial Terrain}
\author{Ana Henrique, Sofia Sheikh, and Benjamin Fields}
\date{}
\begin{document}
\maketitle
\begin{itemize}
    \item \textbf{Topic: 2 Identify Emerging Themes and Technologies}
\end{itemize}

\begin{itemize}
    \item \textbf{Ana Henrique} - SETI Institute Affiliate/ahenrique@setiap.org; \textbf{Sofia Sheikh} - SETI Institute/ssheikh@berkeley.edu; \textbf{Benjamin Fields} - Technosignatures SAG/bnfields72@gmail.com
\end{itemize}

As motivated in the \citet{omnibus} omnibus technosignature DARES response, artificial radio emissions have historically been a major focus of technosignature searches. Looking forward into the next decade, however, the field of technosignatures is excitingly positioned to consider and quantify broader impacts of technology. Considering our advancing capabilities in astronomical instrumentation, we can apply an anthropological lens to expand beyond early technosignature assumptions; e.g., that technological societies will inevitably arrive at radio telescopes.

Throughout human history, the use of tools and techniques to transform the environment has been ubiquitous, albeit with diverse results. Many human societies imprinted permanent changes on the planet's surface as they transformed their surroundings. For purposes that include but are not restricted to agriculture, housing, hunting, burial, and observation of the sky, ``[...] wherever humans have trodden, the natural environment is somehow different, sometimes in barely perceptible ways, sometimes in dramatic ways'' \citep{balee2006time}. On Earth, and possibly beyond, traces of landscape engineering in the form of artificial structures should be clear indicators of technological societies. Such structures constitute a distinct subset of solar system technosignatures, which could also include free-floating or active artifacts --- these are covered in depth in the Lazio et al. DARES RFI submission.

We advocate that the search for technosignatures consider how  societies of the past interacted with their environment, materializing ideas and intentions into the landscape and, consequently, generating nonrandom patterns that can be recognized as artificial. We argue that the landscape could be read as a physical record of intentionality \citep{balee2006time}.  \textbf{Searches for surface technosignatures on rocky solar system bodies are an underexplored and exciting new theme in technosignature work} --- specifically, searches for extraterrestrial artifacts informed by traces of human landscape engineering that have survived for hundreds (sometimes thousands) of years, and that can be identified from above.

Given the ubiquity of humans impacting their planetary surface in long-lasting, technologically mediated ways, it is worth exploring the idea that other terrestrial bodies in the solar system could show similar traces of non-random and intentional patterning. These impacts, much like in the human case, could come from a variety of activities, e.g., habitation, waste disposal, mining, etc. Critically, surface impacts could still be present in planetary bodies that are not currently habitable or planetary bodies that were \textit{never} habitable, as illustrated by human-technological impacts on the moon or the asteroid Dimorphos \citep[the target of the DART mission][]{chabot2024achievement}. In addition, the lack of tectonic or biological activity on these bodies would preserve surface technosignatures for much longer than analogous signatures would remain visible on Earth's surface, potentially for up to hundreds of millions of years \citep{davies2013searching}. 

The probability of intelligent extraterrestrial traces and artifacts on non-Earth solar system objects is difficult to estimate \textit{a priori}. As conjectured by \citet{1975QJRAS..16..128H}, self-replicating interstellar ``Von Neumann'' probes could blanket the entire Milky Way galaxy in a small fraction of our galaxy's total age; \citet{1975QJRAS..16..128H} was pessimistic about the existence of spacefaring civilizations due to a perceived lack of any confirmed remnants from such visitations. However, as deduced by \cite{HAQQMISRA201215}, a majority of surfaces in the solar system have not been explored at sufficient spatial resolution to rule out the presence of extraterrestrial artifacts. As far as we know, humans are the first and only spaceflight-capable species that has evolved in our solar system over its history. However, \citep{Wright_2017} also considers the possibility of technological species which may have evolved in our solar system prior to humans, either on the early Earth, a potentially habitable pre-runaway greenhouse Venus or a young wet Mars. Plate tectonics, weathering and other geological processes are likely to erase such traces on Earth or Venus on relatively short timescales, geologically speaking \citep{schmidt2019silurian}. However, if a prior indigenous technological species originated elsewhere in the solar system or eventually developed spacefaring capabilities, we could still search for traces on geologically dead surfaces; e.g. KBOs, asteroids, rocky moons, or Mars.  

None of these plausibility arguments demonstrate that there are any definite surface technosignatures on solar system bodies. However, they do provide a strong theoretical rationale for why such searches are well-justified. Given the other advantages of such technosignatures (their longevity, the ability to search for them \textit{in situ}, their conceptual breadth), \textbf{setting quantitative limits on surface technosignatures is worthwhile and achievable, and it informs us about the distribution of intelligent life in the Universe}. The study of planetary surface anomalies would also have ancillary benefits for planetary science and geophysics, and could lead to the discovery of novel natural structures in addition to any potential technosignatures. 

Surface technosignature searches represent an emerging technique due to \textbf{two critical developments in the last decade: improved planetary surface datasets and machine learning methods}. Critically, due to these developments, they are ``low risk-high reward'' in terms of their science return: we can use existing databases and archival data to perform novel searches in an efficient, low-cost way. 

Planetary surface imaging datasets have improved immensely in the last decade. The NASA Lunar Reconnaissance Orbiter (LRO) Near-Angle Camera (NAC) has produced petabytes of high spatial resolution surface imagery of the moon, in a dataset of over two million images to date \citep{robinson2010lunar}. The Mars Reconnaissance Orbiter (MRO) HiRISE camera can achieve a resolution of 30 centimeters per pixel in the best case, with more than 92 thousand images in the publicly available catalog\footnote{\url{https://www.uahirise.org/catalog/}}. Search modalities beyond optical imaging, such as remote sensing, can also be leveraged to search for surface technosignatures. LiDAR and ground penetrating radar (GPR) can extend the search to subterranean structures or those buried in e.g., regolith. LiDAR and GPR can also substantiate and reveal additional details about any candidate structures initially identified in optical surface images. Reflectance spectroscopy can be used to search for anomalous artificial materials, which may have a distinct chemical/mineralogical profile compared to the background of natural geological regolith. Using such a multi-modal approach can ensure a scientifically robust search and multiple independent lines of inquiry for any given anomaly.

In order to search for surface technosignatures in these datasets, we can turn to machine learning methods. Unsupervised or self-supervised anomaly detection methods can be used if we expect that the fraction of surface coverage from any surface technosignature is relatively small (a reasonable assumption, as no surface technosignatures, or even candidates, have been discovered to date). \citet{lesnikowski2024automated} recently developed a self-supervised anomaly detection method (which does not require a training set) for LRO NAC data, which successfully returned not just known technosignatures (i.e., the Apollo 17 lander) but also more traditional planetary science objects-of-interest such as recent impacts and volcanic pits. 

Alternately, we can use our prior knowledge and existing datasets of surface technosignatures on the Earth to serve as training sets for supervised anomaly detection or classification methods. Massive terrestrial structures that have withstood the test of time --- such as the Kailasa Temple in Maharashtra, India, the Ziggurat of Ur in Iraq, and the Inca agricultural terraces in Peru (see Figure 1) --- provide examples of non-random patterns produced by Earth societies. They are clear indicators of human intentionality and technological capacity, and have been imaged from orbit by similar technologies to our assets around other planetary bodies. 

The unsupervised and supervised approaches to surface technosignatures each present unique advantages and disadvantages.The supervised method is currently used by archaeologists to find ruins/structures in satellite imagery \citep{app13042613}, so we know its effectiveness at distinguishing artificial structures from surrounding natural geology. It also allows for focused and directed hypothesis testing, rather than a blind approach. However, the unsupervised approach 
mitigates potential anthropocentric biases, which may also reduce the odds of false positives (i.e., pareidolia, spurious misinterpretations of natural formations) or false negatives (i.e., the algorithm ignoring a technosignature because it happens not to match the training set). For this reason, we advocate for development of both supervised and unsupervised methods in parallel---each is likely to yield a different set of results, and thus provide unique and complementary insights.

In looking for traces of surface modifications and non-random patterning as indicators of technology, we would be looking for other versions of ourselves  \citep{cabrol2016alien}. However, we would also be amplifying our perception of what a technosignature can be by considering how Earth societies have employed technology and impacted the surface for thousands of years. Additionally, merging methodologies between terrestrial archaeology and planetary science can yield exciting interdisciplinary collaboration, which has a positive impact on academia writ large.

Despite the clear path for setting upper limits on surface technosignatures using these datasets and algorithms, no quantifications of this sort have yet been made, which was identified as an \textbf{area of opportunity in the NASA Technosignatures Workshop Report \citep{participants2018nasa}}. In the framework of the Nine Axes of Merit, surface technosignatures are extremely meritorious: they can be investigated in a low-cost way with current missions and datasets, they have ancillary science benefits, they are long-lasting, likely to be unambiguous (or at least, have the potential for \textit{in situ} follow-up) and could be information-rich. They also take advantage of advances in machine learning techniques and mission instrumentation. For all of these reasons, \textbf{surface technosignatures reside in an important niche of parameter space and should be represented in a robust technosignature portfolio}.
\begin{figure}
    \centering
    \includegraphics[width=0.75\linewidth]{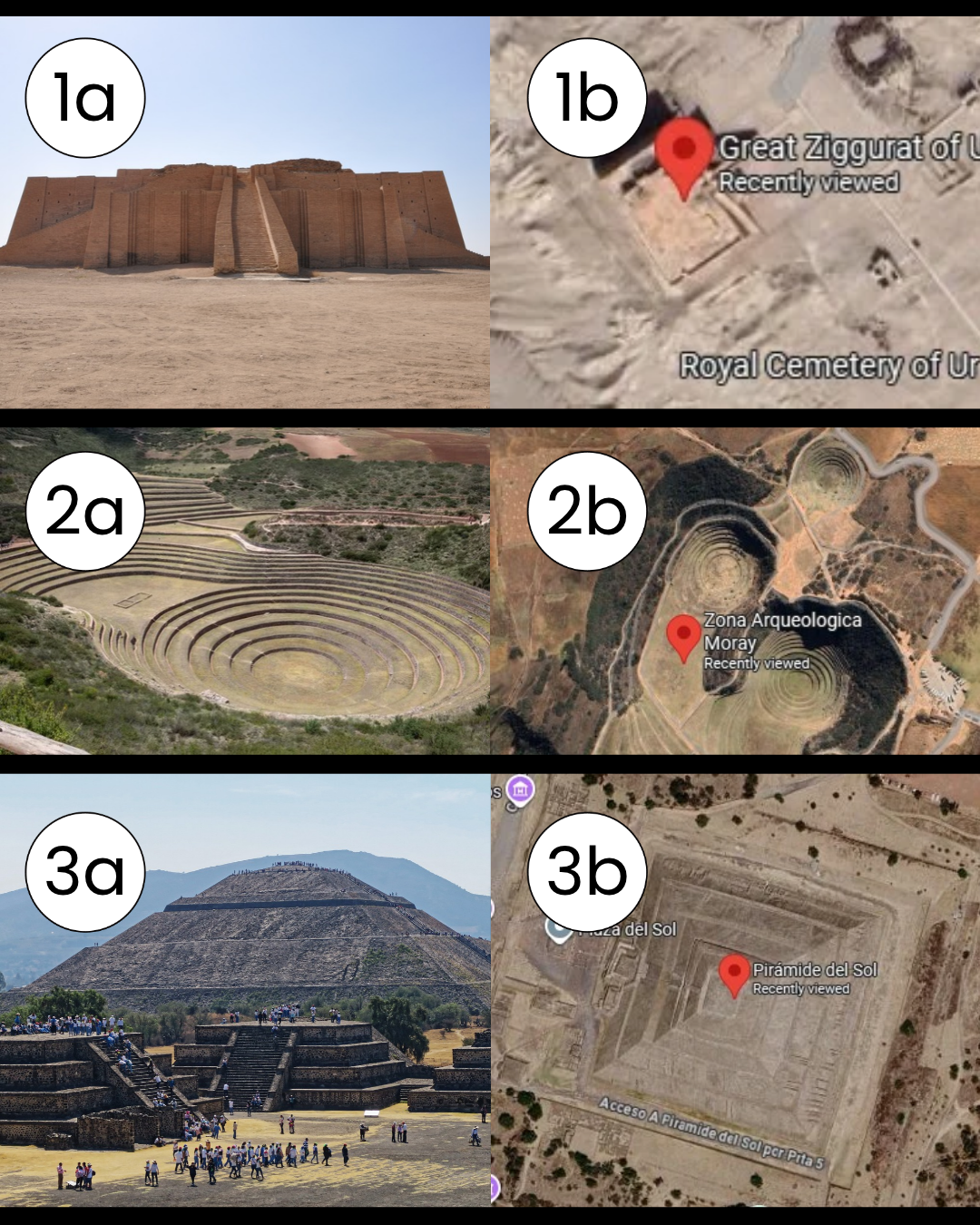}
    \caption{Three examples of surface technosignatures on Earth that are over 500 years old and are visible in orbital satellite imagery: 1) Ziggurat of Ur 2) Inca Agricultural Terraces and 3) Pyramid of the Sun. The a) images show the structures from a near-surface perspective while the b) images are representative of the structures appearance in satellite imagery, as sourced from Google Maps. Two takeaways can be drawn here: surface technosignatures can be readily identified from satellite imagery, and we have a large, consistent training dataset of examples of human surface technosignatures from the Earth itself. Image credits: © Ziggurat of Ur: Qahtan Al-Abeed | ©  Inca Agricultural Terraces:  PsamatheM | ©  Pyramid of the Sun: Burkhard Mücke}
    
    \label{fig:structures_}
\end{figure}

\bibliographystyle{abbrvnat}
\bibliography{references}

\end{document}